\documentclass[12pt]{iopart}
\usepackage{graphicx}
\usepackage{bm}
\usepackage{color}

\newcommand{\calQ}{{\mathcal{Q}}}
\newcommand{\calW}{{\mathcal{W}}}
\newcommand{\calS}{{\mathcal{S}}}
\newcommand{\calY}{{\mathcal{Y}}}
\newcommand{\calE}{{\mathcal{E}}}

\newcommand{\calP}{{\mathcal{P}}}
\newcommand{\calT}{{\mathcal{T}}}

\newcommand{\calD}{{\mathcal{D}}}

\newcommand{\sfF}{{\mathsf{F}}}

\newcommand{\sfA}{{\mathsf{A}}}
\newcommand{\sfS}{{\mathsf{S}}}
\newcommand{\sfI}{{\mathsf{I}}}

\newcommand{\bmf}{{\bm{f}}}
\newcommand{\bmx}{{\bm{x}}}
\newcommand{\bmxi}{{\bm{\xi}}}

\begin{document}

\title{Fluctuations and correlations in nonequilibrium systems} 
\author{Jae Dong Noh}
\address{Department of Physics, University of Seoul, Seoul 130-743,
 Korea}
\address{School of Physics, Korea Institute for Advanced Study,
Seoul 130-722,  Korea}

\date{April 7, 2013}

\begin{abstract}
Nonequilibrium systems exchange the energy with an environment in the form
of work and heat. The work done on a system obeys the fluctuation 
theorem, while the dissipated heat which differs from the work 
by the internal energy change does not. 
We derive the modified fluctuation relation 
for the heat in the overdamped Langevin system.
It shows that mutual correlations among the
work, the heat, and the internal energy change are responsible for the 
different fluctuation property of the work and the heat. The mutual
correlation is investigated in detail 
in a two-dimensional linear diffusion system. 
We develop an analytic method which allows one to calculate the large deviation 
function for the joint probability distributions. 
We find that the heat and the internal energy change have a negative 
correlation, which explains the reason for the breakdown 
of the fluctuation theorem for the heat.
\end{abstract}
\maketitle

\section{Introduction}
Consider a nonequilibrium stochastic system in thermal contact with a heat bath.
It is driven out of equilibrium by a time-dependent perturbation or a
nonconservative force. Due to absence of the detailed balance, a 
nonequilibrium system is characterized by a nonzero probability current
and a net energy flow.
Recently, nonequilibrium fluctuations of thermodynamic quantities such
as a work, an entropy production, and a heat 
have been attracting growing interests. Various
types of fluctuation theorems governing the nonequilibrium fluctuations 
have been found in a thermostated shearing
fluid~\cite{Evans93,Gallavotti95}, systems 
driven by a nonequilibrium work~\cite{Jarzynski97,Crooks99}, 
stochastic Langevin systems~\cite{Kurchan98},
master equation systems~\cite{Lebowitz99}, 
systems in a nonequilibrium steady state~\cite{Hatano01}, 
general stochastic systems~\cite{Seifert05,Esposito10}, 
feedback control systems~\cite{Sagawa10},
systems with odd parity variables~\cite{Spinney12,Lee13}, and so on.
Theoretically, the fluctuation theorems allow us to study the time
irreversibility of nonequilibrium systems. At the same time, they
play an important role in studying small-sized systems 
such as colloidal
particles~\cite{Wang02}, bio-molecules~\cite{Hummer01,Liphardt02}, 
and molecular motors~\cite{Hayashi10}, 
where thermal fluctuation effects are important. 
Further theoretical and
experimental studies are found in recent review
papers~\cite{Ciliberto10,Seifert12}.

We are interested in mutual correlations among the 
thermodynamic quantities. Most studies have been focused on the 
fluctuations of an individual thermodynamic quantity 
with a few exceptions~\cite{Garcia10,Garcia12}. 
The current study is motivated by our recent work on 
the modified fluctuation relation for the heat~\cite{Noh12}. 
The amounts of a work $\calW$ and a heat $\calQ$
during a nonequilibrium process over a time interval $\Delta t$
are constrained by the thermodynamic first law
\begin{equation}\label{1st_law}
\calW = \calQ + \Delta\calE
\end{equation}
where $\Delta\calE$ denotes the change in the internal energy of the system. 
With nonzero net energy flow, both $\calW$ and $\calQ$ scale
with the time interval $\Delta t$ while their difference 
$\Delta\calE$ does not on average. 
Hence, one may expect that the heat would follow 
the same fluctuation theorem obeyed by the work in the large $\Delta t$ limit.
However, various studies found that the fluctuation theorem breaks down for
the heat~\cite{Farago02,Zon03,Zon04,Garnier05,Visco06,Harris06,Rakos08,
Fogedby11,Nemoto12}. The breakdown suggests that the effect of the 
the boundary term $\Delta\calE$ may persist even in the infinite time
limit~\cite{Puglisi06,JLee13}. 
Moreover, the modified fluctuation relation for the
heat derived in Ref.~\cite{Noh12} suggests that the correlation between the
heat and the energy change plays an important role in nonequilibrium
processes. 

Based on this motivation, we investigate the mutual correlations among the
thermodynamic quantities in a linear diffusion system. 
It is simple, but exhibits various
nontrivial genuine nonequilibrium phenomena~\cite{Kwon11,Noh13,Kwon13}.  
We develop a path integral formalism to study the joint probability
distribution analytically. Our study gives a hint on the reason why 
the fluctuations of the work and the heat are different.

This paper is organized as follows: In Sec.~\ref{sec2}, we introduce an
overdamped Langevin dynamics for a nonequilibrium system driven by both a
time-dependent perturbation and a nonconservative force, and review
the stochastic thermodynamics formalism for the fluctuation 
theorem. In Sec.~\ref{sec3}, we derive the
modified fluctuation relation for the heat in a general setting. In
Ref.~\cite{Noh12}, we only considered a time-independent 
nonconservative force. We extend the formalism to include a
time-dependent driving force. In Sec.~\ref{sec4}, we investigate the fluctuations
and the correlations in a two-dimensional linear diffusion system in detail.
We develop an analytic method in \ref{appendix}, 
which allows us to calculate the 
joint probability distributions of the thermodynamic quantities analytically.
We find that there is a strong negative correlation between
$\calQ$ and $\Delta\calE$. 
We conclude the paper with summary in Sec.~\ref{sec5}.

\section{Stochastic Thermodynamics for Langevin systems}\label{sec2}
Consider an overdamped dynamics of a nonequilibrium system in thermal
contact with a heat bath at temperature $T$. When there are $N$ degrees 
of freedom, a configuration is described by an $N$-dimensional column vector
$\bmx=(x_1,\cdots,x_N)^T$. The Langevin equation reads as 
\begin{equation}\label{Leq}
\dot{\bmx}(t) = \bmf + \bmxi(t) \ , 
\end{equation}
where $\bmf$ is a force and $\bmxi=(\xi_1,\cdots,\xi_N)^T$ is a thermal noise 
satisfying
\begin{equation}
\langle \xi_i(t)\rangle = 0 \ , \ \langle \xi_i(t)\xi_j(t')\rangle = 2 
\beta^{-1} \delta_{ij}\delta(t-t') ,
\end{equation}
with the inverse temperature $\beta \equiv 1/T$. The superscript ${}^T$
denotes the transpose. The damping
coefficient and the Boltzmann constant are set to unity.
In general, the force $\bmf = \bmf(\bmx,\alpha)$ is a sum of conservative 
and nonconservative forces as
\begin{equation}
\bmf(\bmx,\alpha) = -{\bm \nabla}_\bmx V(\bmx,\alpha) + \bmf_{nc}(\bmx)
\end{equation}
with an energy function $V(\bmx,\alpha)$. 
The energy function may depend on a time-dependent external parameter 
$\alpha=\alpha(t)$, called a protocol.
The nonconservative force $\bmf_{nc}$ cannot be written as a gradient of 
any scalar function.

When $\bmf_{nc}=0$ and $\alpha$ is independent of $t$, 
the system relaxes into a thermal
equilibrium state characterized by the Boltzmann
distribution~\cite{gardiner,risken}
\begin{equation}\label{P_eq}
P_{eq}(\bmx,\alpha) = \exp[ \beta F(\alpha) - \beta V(\bmx,\alpha) ] 
\end{equation}
with the free energy given by
\begin{equation}\label{free_energy}
F(\alpha) = -\beta^{-1} \ln\left[ \int d\bmx\ e^{-\beta V(\bmx,\alpha)}\right] .
\end{equation}
A nonzero $\bmf_{nc}$ or a time-dependent protocol
$\alpha(t)$ drives the system out of equilibrium.

During the time evolution, the system exchanges the energy with the
environment. Suppose that the system evolves in time along a stochastic path 
or trajectory $\bmx(t)$ during the time interval $t_i\leq t\leq t_f$. 
Then, the nonequilibrium work done on the system is given
by~\cite{Kwon11,Sekimoto98}
\begin{equation}\label{W_def}
\calW[\bmx(t)] = \int_{t_i}^{t_f} dt \left[ \frac{d\alpha}{dt} \frac{\partial
V(\bmx,\alpha)}{\partial \alpha} + \dot\bmx^T \cdot \bmf_{nc}(\bmx) \right] ,
\end{equation}
where the first term is the thermodynamic work required to change
the external parameter and the second term is the work done by the
nonconservative force. The change in the internal energy is given by
\begin{equation}\label{Ed_def}
\Delta \calE[\bmx(t)] = V(\bmx(t_f),\alpha(t_f)) - V(\bmx(t_i),\alpha(t_i)) .
\end{equation}
It can be rewritten as
\begin{eqnarray}
\Delta \calE[\bmx(t)] &=& 
\int_{t_i}^{t_f} dt \ \frac{d}{dt}V(\bmx(t),\alpha(t)) \nonumber \\
& = & \int_{t_i}^{t_f} dt \ \left[ \frac{d\alpha}{dt}\frac{\partial V}{\partial
\alpha} + \dot\bmx^T \cdot \bm\nabla_\bmx V\right] . \label{E_def}
\end{eqnarray}
The heat dissipated into the bath is then given by
\begin{equation}\label{Q_def}
\calQ[\bmx(t)] = \calW[\bmx(t)] - \Delta \calE[\bmx(t)] =
\int_{t_i}^{t_f} dt\ \dot\bmx^T \cdot \bmf .
\end{equation}
Note that the integral in (\ref{W_def}), (\ref{E_def}),
and (\ref{Q_def}) is the Stratonovich integral~\cite{gardiner}.

Since the system follows the stochastic Langevin dynamics, 
thermodynamic quantities $\calW$, $\Delta \calE$, and $\calQ$ 
are random variables.
Stochastic thermodynamics predicts several interesting identities, known as
the fluctuation theorems, for their probability density functions~(PDFs). 
We present a brief summary on the fluctuation theorems~\cite{Seifert12}.

The key feature behind the fluctuation theorem is that the thermodynamic quantity can be written
in terms of a relative entropy~\cite{Cover06} 
between the path probabilities in forward~(F) and reverse~(R) processes. 
The F process is specified by a protocol $\alpha(t)$, a nonconservative 
force $\bmf_{nc}$, and a PDF $P_i$ for the initial configuration.
The R process, corresponding to the F process, is specified by the 
time-reversed protocol $\alpha^{R}(t)\equiv \alpha(t_f-(t-t_i))$, the same
nonconservative force $\bmf_{nc}$, and a PDF $P^R_i$ for the initial
configuration. Consider a following quantity associated with a path
$\bmx(t_i\leq t \leq t_f)$
\begin{equation}\label{Y_def}
\calY[\bmx(t)] \equiv \ln \left(
\frac{\calP[\bmx(t)]}{\calP^R[\bmx^R(t)]} \right) \ ,
\end{equation}
where $\calP[\bmx(t)]$ and $\calP^{R}[\bmx(t)]$ are the
probability distribution of the system following a trajectory $\bmx(t)$ 
in the F and R processes, respectively, and $\bmx^R(t)\equiv
\bmx(t_f-(t-t_i))$ is the time-reversed trajectory. 
Such a quantity satisfies the identity
\begin{equation}\label{IFT}
\left\langle e^{-\calY[\bmx(t)]} \right\rangle =1 \ ,
\end{equation}
where the average is taken over all paths $\bmx(t)$ in the F process.
The Jensen's inequality then yields that
\begin{equation}\label{2ndlaw}
\langle \calY[\bmx(t)] \rangle \geq 0 \ .
\end{equation}

The path probability distribution is given by 
\begin{equation}
\calP^{(R)}[\bmx(t)] = \calT^{(R)}[\bmx(t)|\bmx(t_i)] P^{(R)}_i(\bmx_i)
\end{equation}
with the conditional probability $\calT^{(R)}[\bmx(t)|\bmx(t_i)]$ for a
path $\bmx(t)$ starting at $\bmx(t=t_i)$ 
in the F~(without superscript) and 
R~(with superscript ${}^R$) processes. Using the Onsager-Machlup
formalism~\cite{Onsager53},
one can show that the ratio between the conditional probabilities is equal to
the heat dissipation~\cite{Kurchan98,Seifert05,Kwon11}:
\begin{equation}\label{micro_rev}
\calQ[\bmx(t)] = \beta^{-1} \ln\left(
\frac{\calT[\bmx(t)|\bmx_i]}
     {\calT^R[\bmx^R(t)|\bmx_f]}\right) 
\end{equation}
with $\bmx_i=\bmx(t=t_i)$ and $\bmx_f = \bmx^R(t=t_i) = \bmx(t_f)$.
Through this relation, $\calY$ can represent a thermodynamic quantity 
with a suitable choice of $P_i$ and $P^R_i$.
Such a quantity then automatically satisfies the identity of (\ref{IFT}),
which is referred to as an integral fluctuation theorem~(IFT).

One can obtain the well-known ITF for the total entropy change 
\begin{equation}
\Delta\calS=
-\ln\left[\frac{P_f(\bmx(t_f))}{P_i(\bmx(t_i))}\right] + \beta\calQ
\end{equation}
by choosing $P^R_i(\bmx)$ as the PDF $P_f(\bmx)$ of finding the system,
evolving from the initial PDF $P_i(\bmx)$ at $t=t_i$ in the F
process, in configuration $\bmx$ at time $t=t_f$.  The first term
corresponds to the change in the Shannon entropy of the system
$\Delta\calS_{sys}$, while the second term 
the change of the heat bath entropy. 

One can also obtain the ITF for the nonequilibrium work, known as the
Jarzynski equality~\cite{Jarzynski97}, 
by choosing $P_i(\bmx) = P_{eq}(\bmx,\alpha_i=\alpha(t_i))$ and
$P_i^R(\bmx) = P_{eq}(\bmx,\alpha_f=\alpha(t_f))$ 
with the Boltzmann distribution in
(\ref{P_eq}). With this choice, one finds that $\calY =
\beta(\calQ+\Delta\calE-\Delta F)= \beta(\calW-\Delta F)$ with the free
energy difference $\Delta F \equiv F(\alpha_f)-F(\alpha_i)$.

In a certain circumstance, the functional
$\calY[\bmx(t)]$ and the functional 
$\calY^R[\bmx(t)] \equiv \ln[\calP^R[\bmx(t)]/\calP^{RR}[\bmx^R(t)]]$ may
satisfy a relation $\calY^R[\bmx^R(t)]=-\calY[\bmx(t)]$. It holds only when
the reverse process of the R process is equivalent to the original F
process, i.e., $\calP^{RR}[\bmx(t)]=\calP[\bmx(t)]$, which is called an
involution property~\cite{Garcia12}. Then, the PDF $P(Y) \equiv \langle
\delta(\calY[\bmx(t)]-Y)\rangle$ for the F process and the PDF $P_R(Y)
\equiv \langle \delta(\calY^R[\bmx(t)]-Y)\rangle_R$ for the R process satisfy 
the identity~\cite{Seifert05}
\begin{equation}\label{DFT}
\frac{P(Y)}{P_R(-Y)} = e^Y \ ,
\end{equation}
which is called a detailed fluctuation theorem~(DFT) in comparison to the
IFT. The DFT implies the IFT, while the converse is not true in general.
Using (\ref{Y_def}) and the involution property, 
one can derive (\ref{DFT}) as follows:
\begin{eqnarray*}
P(Y) &=& \int[{D}\bmx] \delta(\calY[\bmx]-Y) \calP[\bmx] \\
     &=& \int[{D}\bmx] \delta(\calY[\bmx]-Y) \calP^R[\bmx^R] 
         e^{\calY[\bmx]} \\
     &=& \int[{D}\bmx^R] \delta(-\calY^R[\bmx^R]-Y) \calP^R[\bmx^R] 
         e^{-\calY^R[\bmx^R]} \\
     &=& e^Y P_R(-Y)  \ ,
\end{eqnarray*}
where $\int[{D}\bmx]$ denotes the path integral.

The choice of $P_i(\bmx)=P_{eq}(\bmx,\alpha_i)$ and
$P_i^R(\bmx)=P_{eq}(\bmx,\alpha_f)$ leading to $\calY = \beta(\calW-\Delta F)$
preserves the involution property. 
Hence, $\beta(\calW-\Delta F)$ satisfies the DFT,
which is referred to as the Crooks fluctuation theorem~\cite{Crooks99}.

The total entropy production $\Delta\calS$ corresponds to the choice of
$P_i^R(\bmx)=P_f(\bmx)$. In general nonequilibrium systems, 
the PDF $P_f(\bmx)$ does return to the original PDF $P_i(\bmx)$ 
under the R process. Hence, the total entropy production does
not obey the DFT. There is an exceptional case. 
Suppose that the external parameter $\alpha$ is a time-independent
constant and the system is in the nonequilibrium steady state
initially. Then, the system remains at the steady
state at all times and the involution property holds. Hence, the entropy
production in the nonequilibrium steady state obeys the DFT.

Note that the functional in (\ref{Y_def}) is the relative
entropy of two probability distributions $\calP$ and $\calP^R$.
One may adopt the path probability distribution from 
a different dynamics in (\ref{Y_def}) instead of  the R process.
For example, if one chooses the so-called dual or adjoint dynamics, then 
the heat can be decomposed into two parts as $\calQ = \calQ_{ex} +
\calQ_{hk}$~\cite{Hatano01,Esposito10}. When a system is in a nonequilibrium
steady state, it gains a work done by $\bmf_{nc}$ and dissipates the same
amount of a heat into the heat bath. 
The house-keeping heat $\calQ_{hk}$ refers to the heat necessary to
maintain the nonequilibrium steady state. The excess heat $\calQ_{ex}$ 
refers to the heat dissipated in transient
dynamics~\cite{Hatano01}.
As well as the total entropy production $\Delta \calS = \Delta \calS_{sys} +
\beta \calQ_{ex}+\beta \calQ_{hk}$, each of 
$\Delta\calS_{sys}+\beta \calQ_{ex}$ and $\beta\Delta\calQ_{hk}$ is known to
satisfy the IFT, respectively~\cite{Esposito10}. The total heat may be
decomposed in a different way for underdamped systems~\cite{Spinney12,Lee13}, 
which is not covered in the paper.

\section{Modified fluctuation relation for heat}\label{sec3}
It is an interesting question whether the heat also satisfies the 
fluctuation theorem. A careful consideration of
(\ref{Y_def}) provides an example with the affirmative
answer. If one chooses the uniform distribution for $P_i$ and $P^R_i$,
the functional $\calY$ becomes equal to $\beta\calQ$. 
In this case, the heat can satisfy the fluctuation theorem. 
However, this is a peculiar case since the uniform distribution 
cannot be realized in systems whose phase space is unbounded.
So the main question is whether the heat satisfies the fluctuation theorem 
in systems following the equilibrium Boltzmann distribution initially.

Theoretical and experimental studies have shown that the 
heat, unlike the work, does not obey the fluctuation theorem even in the
infinite $\Delta t = t_f-t_i$ limit~\cite{Zon03,Zon04,Garnier05}. 
It is rather surprising 
because the work and the heat are proportional to $\Delta t$ on
average while their difference $\langle\calW\rangle - \langle\calQ\rangle
=\langle\Delta\calE\rangle$ does not scale with $\Delta t$. The breakdown is
attributed to a rare but large fluctuation of the system in the 
energy landscape~\cite{Nemoto12}.

Recently, a general relation, called the modified fluctuation relation, 
was found for the heat~\cite{Noh12}. 
It was derived for systems only with a nonconservative force. 
Here we generalize it to cover the systems driven by both 
a time-dependent protocol $\alpha(t)$ and a nonconservative force.

The system starts from the Boltzmann distribution of (\ref{P_eq}) with
$\alpha=\alpha_i~(\alpha_f)$ in the F~(R) process so that the functional 
in (\ref{Y_def}) becomes as $\calY[\bmx] = \ln({\calP[\bmx]}/{\calP^R[\bmx^R]}) =
\beta(\calQ[\bmx]+\Delta\calE[\bmx]-\Delta F)$. We rewrite it as a relation
between the path probabilities in the F and R processes: 
\begin{equation}\label{PfPr_he}
\calP[\bmx] = e^{\beta(\calQ[\bmx]+\Delta\calE[\bmx]-\Delta F)} \ 
\calP^R[\bmx^R] \ .
\end{equation}
Multiplying both sides with
$\delta(\calQ[\bmx]-Q)\delta(\Delta\calE[\bmx]-E)$ and integrating over all
paths, we obtain the fluctuation theorem
\begin{equation}\label{FT_he}
P_{\rm he}(Q,E) = e^{\beta(Q+E-\Delta F)} {P^R_{\rm he}(-Q,-E)}\ .
\end{equation}
for the joint PDF of the heat and the energy change 
defined as
\begin{equation}\label{P_he}
P_{\rm he}(Q,E) \equiv \langle \delta(\calQ[\bmx]-Q) \delta(
\Delta \calE[\bmx]- E) \rangle \ .
\end{equation}
In deriving (\ref{FT_he}), we have used that $\calQ^R[\bmx^R]=-\calQ[\bmx]$,
$\Delta\calE^R[\bmx^R] = - \Delta\calE[\bmx]$, and $\int[{D}\bmx] =
\int[{D}\bmx^R]$. The fluctuation theorem for the joint PDF 
was also considered in Ref.~\cite{Garcia10,Garcia12}.

Hereafter, we will use `h', `w', and `e' in the subscript for a PDF of a
heat, work, and energy change, respectively. When there are multiple
subscripts as in (\ref{P_he}), it should be understood as a joint PDF of
corresponding quantities. In addition, we will use a `$|$' 
in the subscript for a conditional PDF. 
For example, $P_{\rm e|h}(E|Q)$ denotes the conditional probability that the
energy change is $E$ given that the heat dissipation is $Q$.

The marginal distribution of the heat is given by $P_{\rm h}(Q)=\int dE
P_{\rm he}(Q,E)$. Integrating both sides of (\ref{FT_he}) over $E$, 
we obtain that
\begin{equation}
P_{\rm h}(Q) = e^{\beta(Q-\Delta F)} \left( \int dE P^R_{\rm he}(-Q,-E)
e^{\beta E}\right)
\ .
\end{equation}
The term in the parenthesis is simplified by using the conditional
probability $P_{\rm e|h}(E|Q) = P_{\rm he}(Q,E) / P_{\rm h}(Q)$.
It leads to the relation
\begin{equation}\label{FTp_Q}
\frac{P_{\rm h}(Q)}{P_{\rm h}^R(-Q)} = e^{\beta(Q-\Delta F)} \Psi^R(-Q) \ .
\end{equation}
where
\begin{equation}
\Psi^R(Q) \equiv \int dE e^{-\beta E} P^R_{\rm e|h}(E|Q) \ .
\end{equation}
Applying the relation (\ref{FTp_Q}) to the R process and using the involution 
property that the reverse process of the R process is equivalent to the 
F process, one can show that 
\begin{equation}
\Psi^R(-Q) = 1 / \Psi(Q)
\end{equation}
where
\begin{equation}\label{Psi_def}
\Psi(Q) \equiv \int dE e^{-\beta E} P_{\rm e|h}(E|Q) \ .
\end{equation}
Therefore, we obtain the modified fluctuation relation for the heat
\begin{equation}\label{DFT_Q}
\frac{P_{\rm h}(Q)}{P_{\rm h}^R(-Q)} = e^{\beta(Q-\Delta F)} / \Psi(Q) \ .
\end{equation}

Being compared with the DFT in (\ref{DFT}), the modified fluctuation
relation 
is dressed by the additional factor $1/\Psi(Q)=\Psi^R(-Q)$. In order to
check whether the heat satisfies the IFT in (\ref{IFT}), we evaluate $\langle
e^{-\beta\calQ[\bmx]}\rangle$ using (\ref{PfPr_he}) and (\ref{FT_he}). It
yields that
\begin{equation}
\left\langle e^{-\beta(\calQ[\bmx]-\Delta F)} \right\rangle = 
\left\langle e^{-\beta \Delta\calE^R[\bmx]}\right\rangle_R \ .
\end{equation}
It is also dressed by the additional factor representing the fluctuation of
the energy change in the R process. 

The additional factor $\Psi(Q)$ in (\ref{DFT_Q}) reflects a correlation between 
the heat and the energy change during a nonequilibrium process. This calls 
for the study of mutual correlations between thermodynamic quantities as
well as their individual fluctuations. 
The energy change, the work, and the heat are
constrained by the thermodynamic first law in (\ref{1st_law}).
So, it suffices to specify a single joint PDF, e.g., $P_{\rm we}(W,E)$.
The other joint PDFs are given by 
\begin{equation}\label{he_we}
P_{\rm he}(Q,E) = P_{\rm we}(Q+E,E)
\end{equation}
and
\begin{equation}
P_{\rm wh}(W,Q) = P_{\rm we}(W,W-Q) \ ,
\end{equation}
from which the marginal distributions
$P_{\rm w}(W)$, $P_{\rm h}(Q)$, and $P_{\rm e}(E)$ are obtained.

It is convenient to deal with the moment generating 
functions~(MGFs). We will use the symbol $G$ to denote a MGF, 
and adopt the same subscript notation for the MGF as that for the PDF. 
Parameters $\lambda$, $\kappa$, and $\eta$ will be used as the
conjugate variables for the work, the energy change, and the heat,
respectively. For example, $G_{\rm we}(\lambda,\kappa)$ stands for the MGF 
defined as
\begin{equation}
G_{\rm we}(\lambda,\kappa) \equiv \int dQ\int dE P_{\rm we}(W,E)
e^{-\beta(\lambda W+\kappa E)} \ .
\end{equation}
It is easy to show that the other MGFs for the joint PDFs are given by
\begin{equation}\label{G_he}
G_{\rm he}(\eta,\kappa) = G_{\rm we}(\eta,-\eta+\kappa)
\end{equation}
and 
\begin{equation}
G_{\rm wh}(\lambda,\eta) = G_{\rm we}(\lambda+\eta,-\eta) \  . 
\end{equation}
The MGFs for the marginal distributions are given by 
$G_{\rm w}(\lambda)=G_{\rm we}(\lambda,0)$, $G_{\rm e}(\kappa)=G_{\rm
we}(0,\kappa)$, and $G_{\rm h}(\eta) = G_{\rm we}(\eta,-\eta)$.

In terms of the MGF, the DFT or the Crooks fluctuation theorem for the work 
can be written as
\begin{equation}\label{FTG_W}
G_{\rm w}(\lambda) = e^{-\beta \Delta F}  G^R_{\rm w}(1-\lambda)  \ .
\end{equation}
or  
\begin{equation}\label{FTG_W2}
G_{\rm we}(\lambda,0) = e^{-\beta\Delta F} G^R_{\rm we}(1-\lambda,0) \ .
\end{equation}
The modified fluctuation relation for the heat 
in (\ref{DFT_Q}) can be rewritten as
\begin{equation}\label{FTG_Q}
G_{\rm he}(\eta,0) = e^{-\beta \Delta F} G^R_{\rm he}(1-\eta,1)  \ .
\end{equation}
In comparison with (\ref{FTG_W2}), the heat fluctuation deviates from the DFT
by the factor ${G^R_{\rm he}(1-\eta,1)}/{G^R_{\rm he}(1-\eta,0)}$.

\section{Linear diffusion system}\label{sec4}
In this section, we investigate an exactly solvable 
two-dimensional linear diffusion system. Fluctuations and correlations are
studied in detail to demonstrate the fluctuation theorems for the joint PDF
and the modified fluctuation relation for the heat. Especially, we focus
on the question whether the DFT for the heat holds asymptotically
in the infinite time-interval limit or not. 

We investigate the nonequilibrium fluctuations in a system 
exerted by a linear force~\cite{Kwon11,Noh13}
\begin{equation}\label{F_lin}
\bmf(\bmx) = -\sfF \cdot \bmx
\end{equation}
with a time-independent force matrix $\sfF$. The force is given by the 
sum of a conservative force $\bmf_c(\bmx) = -\sfF_s \cdot \bmx$ and
a nonconservative force $\bmf_{nc}(\bmx) = -\sfF_a \cdot \bmx$ 
with the symmetric component $\sfF_s = (\sfF+\sfF^T)/2$ and the anti-symmetric 
component $\sfF_a = (\sfF-\sfF^T)/2$~\cite{comment1}. 
The conservative force is also written as 
$\bmf_{c}(\bmx)=-\sfF_s \cdot \bmx = -\bm{\nabla}_\bmx V(\bmx)$ with the
energy function $V(\bmx) \equiv \frac{1}{2} \bmx^T \cdot \sfF_s \cdot \bmx$.
The work, heat, and energy change for a path $\bmx(t_i\leq t\leq t_f)$
are given by
\begin{eqnarray}
\calW[\bmx] &=& 
   - \int_{t_i}^{t_f} dt\ \dot{\bmx}^T(t) \cdot \sfF_a \cdot \bmx(t)
     \label{LDS_W} \\
\calQ[\bmx] &=& 
   - \int_{t_i}^{t_f} dt\ \dot{\bmx}^T(t) \cdot \sfF \cdot \bmx(t) \\
\Delta\calE[\bmx] &=& \frac{1}{2}\bmx^T_f \cdot \sfF_s \cdot \bmx_f - 
                      \frac{1}{2}\bmx^T_i \cdot \sfF_s \cdot \bmx_i  
\end{eqnarray}
where the integral is of Stratonovich type, 
$\bmx_i = \bmx(t_i)$, and $\bmx_f=\bmx(t_f)$. 
Without time-dependent driving force, the F and R processes are the same
and $\Delta F =0$. Hereafter, the energy will be measured in unit of 
$k_B T$ and the temperature be set to unity. The system is assumed to be 
in the equilibrium state with the Boltzmann distribution 
given in (\ref{P_eq}) initially.

The linear system covers a wide range of physical systems such as a colloidal
particle trapped in an optical tweezer~\cite{Wang02,Zon03,Zon04}, harmonic
networks~\cite{Kundu11,Saito11,Sabhapandit12,Fogedby12}, and
RC circuits~\cite{Garnier05,Ciliberto13}.
In a linear diffusion system~\cite{Kwon11,Noh13}, 
the equilibrium relaxation dynamics and the
nonequilibrium driven dynamics are competing with each other. The competition
leads to an intriguing dynamical behavior.
For instance, a two-dimensional linear diffusion system
undergoes multiple locking-unlocking dynamical transitions with time 
in the tail shape of $P_{\rm w}(W)$~\cite{Noh13}.

In this work, we consider a simplest force matrix in two dimensions:
\begin{equation}\label{F_solvable}
\sfF = \left( \begin{array}{cc}
1 & \varepsilon \\
-\varepsilon & 1 
\end{array} \right)  = \sfI + i \varepsilon \sigma_y
\end{equation}
with the identity matrix $\sfI$ and the $y$ component of the Pauli matrix
$\sigma_y$.
With this force matrix, the two-dimensional linear diffusion system 
describes a particle trapped in an isotropic harmonic potential
$V(\bmx) = (x_1^2+x_2^2)/2$ and driven by a swirling
nonconservative force $\bmf_{nc} = -i\varepsilon \sigma_y \cdot \bmx$ 
of strength
$\varepsilon$. This system does not exhibit the locking-unlocking transition
observed in a system with an anisotropic harmonic potential~\cite{Noh13}. 
Yet, it displays nontrivial correlations between thermodynamic
quantities.

In a linear diffusion system, the work fluctuation can be studied by using
the path integral formalism developed in Ref.~\cite{Kwon11}. 
We generalize the formalism to cover the joint PDF, which is explained in~\ref{appendix}.
The formal solution for the MGF $G_{\rm we}(\lambda,\kappa)$ is given 
in (\ref{Gwe_formal}). To obtain the explicit solution, one needs to solve the
differential equation in (\ref{dAdt}) for $\tilde\sfA(t)$ with the initial
condition
\begin{equation}\label{A_ini}
\tilde\sfA(t=t_i) = (1-\kappa) \sfI  \ .
\end{equation}
For convenience, we will set $t_i=0$ hereafter.

\begin{figure}[t]
\includegraphics*[width=\columnwidth]{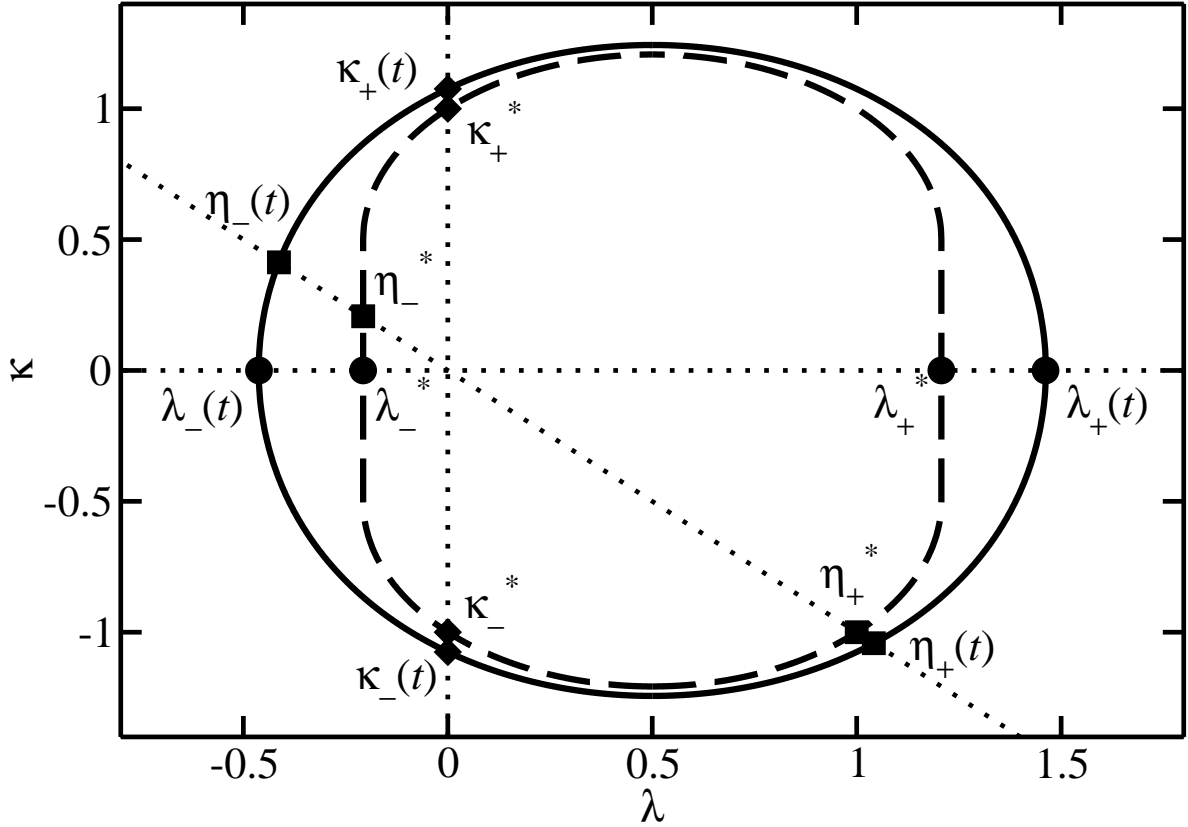}
\caption{Line of singularity of $B(\lambda,\kappa)$ at $t=1$~(solid line) 
and $t=\infty$~(dashed line).}\label{fig1}
\end{figure}

The auxiliary matrices in (\ref{dAdt}) is given by $\tilde\sfF =
\sfI+i\varepsilon(1-2\lambda)\sigma_y$ and $\mathsf{\Lambda}=2\varepsilon^2
\lambda(1-\lambda)\sfI$. The off-diagonal elements of $\tilde{\sfF}$
are anti-symmetric and $\mathsf{\Lambda}\propto\sfI$. 
Hence, $\tilde{\sfA}(t)$, starting from the initial condition in 
(\ref{A_ini}), is proportional to the identity matrix at all $t$. 
Substituting $\tilde\sfA(t)$ with  $z(t) \sfI$ in (\ref{dAdt}), 
one obtains a differential equation for $z(t)$:
\begin{equation}
\dot z  = - 2 z^2 + 2 z + 2 \varepsilon^2 \lambda(1-\lambda)
\end{equation}
with $z(0) = (1-\kappa)$.
The solution is given by
\begin{equation}
z(t) = \frac{1}{2}\left( 1  + \frac{ (1-2\kappa) + \Omega(\lambda) 
\tanh(\Omega(\lambda)t)}{ 1+ \frac{(1-2\kappa)}{\Omega(\lambda)} 
\tanh(\Omega(\lambda)t)} \right) 
\end{equation}
with
\begin{equation}
\Omega(\lambda) = \sqrt{1-4\varepsilon^2 \lambda(\lambda-1)} \ .
\end{equation}
Inserting the solution into (\ref{Gwe_formal}), one obtains
\begin{equation}\label{Gwe_lin_sol}
G_{\rm{we}}(\lambda,\kappa) = B(\lambda,\kappa) \ ,
\end{equation}
where the function $B(x,y)$ is defined as
\begin{equation}\label{B_def}
B(x,y) \equiv
\frac{e^t}{\cosh(\Omega(x)t)+\frac{(1-4y^2)+\Omega(x)^2}{2\Omega(x)}
\sinh(\Omega(x)t)} \ .
\end{equation}
It is symmetric under $x\to 1-x$ and $y\to -y$.
Note that it diverges when the denominator vanishes. 
We plot the lines of singularity at $t=1$~(solid line) and 
$t=\infty$~(dashed line) in the $\lambda-\kappa$ plane in Fig.~\ref{fig1}. 
The solid line is obtained numerically. 
Taking $t\to\infty$ limit in (\ref{B_def}), one finds that 
the dashed line is parameterized as $(\Omega(\lambda)-1)^2=4\kappa^2$. 
It consists of the curved segments
$\kappa = \pm (\Omega(\lambda)-1)/2$
for $\lambda_-^* \leq \lambda \leq \lambda_+^*$ and 
$ |\kappa| \geq 1/2$, and the linear segments
$\lambda = \lambda_{\pm}^*$ for $|\kappa|<1/2$, where
\begin{equation}\label{lambda_pm}
\lambda_{\pm}^* = \frac{1}{2}\pm
\frac{\sqrt{1+\varepsilon^2}}{2\varepsilon}\ .
\end{equation}
The singularity provides an information on the asymptotic tail behavior of
the probability distribution~\cite{Farago02,Visco06}, which will be analyzed
further. 

\subsection{Fluctuations of work}
The MGF of the work is given by 
\begin{equation}\label{Gw_solution}
G_{\rm{w}}(\lambda) = G_{\rm we}(\lambda,0) = B(\lambda,0) \ .
\end{equation}
The symmetry property of the function $B$ ensures the DFT 
in (\ref{FTG_W}) with $\Delta F=0$. 
The mean value of the work is given by
\begin{equation}
\langle \calW\rangle = \left. 
-\frac{dG_{\rm{w}}(\lambda)}{d\lambda}\right|_{\lambda=0} =
2\varepsilon^2 t \ .
\end{equation}
It grows linearly with time $t$. That is, the driving force
$\bmf_{nc}$ performs a work at the uniform rate $2\varepsilon^2$.

At a given value of $t$, $G_{\rm w}(\lambda)$ has simple poles at 
$\lambda_{+}(t)>1$ and $\lambda_{-}(t)<0$, which are marked with 
the closed circles in Fig.~\ref{fig1}.
Due to the symmetry $G_{\rm w}(\lambda) = G_{\rm w}(1-\lambda)$, 
$\lambda_+(t) + \lambda_-(t) = 1$.
The simple poles indicate exponential tails in $P_{\rm w}(W)$:
\begin{equation}\label{tail_Pw}
P_{\rm w}(W) \sim \left\{ \begin{array}{ccc}
e^{\lambda_+(t)W} &,& W \ll -1 \ , \\ [2mm]
e^{\lambda_-(t)W} &,& W \gg 1  \ .
\end{array}\right.
\end{equation}
The tail shape of $P_{\rm w}(W)$ can be characterized more quantitatively 
by the large deviation functions~(LDFs)~\cite{Touchette09} defined as
\begin{eqnarray}
\pi_{\rm w}(w) &\equiv& - \lim_{t\to\infty} \frac{1}{t}\ln
P_{\rm w}(W=wt) \ , \label{e_w}  \\
\gamma_{{\rm w}}(\lambda) &\equiv& - \lim_{t\to\infty} \frac{1}{t}\ln
G_{{\rm w}}(\lambda) \ . \label{e_hatw} 
\end{eqnarray}
The symbol $\pi~(\gamma)$ is used to denote the LDF for the PDF~(MGF).
They are related through the Legendre transformation
\begin{equation}\label{W_leg}
\pi_{\rm w}(w) = \max_{\lambda}\{ \gamma_{{\rm w}}(\lambda) - \lambda w\} \ .
\end{equation}

By taking the $t\to\infty$ limit of (\ref{Gw_solution}), we obtain that
\begin{equation}\label{ldf_gw}
\gamma_{{\rm w}}(\lambda) = \left\{ \begin{array}{ccc}
\Omega(\lambda) - 1 &,& \lambda_-^* < \lambda < \lambda_+^* \\ [2mm]
-\infty &,& \mbox{otherwise}\end{array}\right. \ .
\end{equation}
So the Legendre transformation yields that
\begin{equation}\label{ldf_pw}
\pi_{\rm w}(w) =
\sqrt{\frac{(1+\varepsilon^2)(w^2+4\varepsilon^2)}{4\varepsilon^2}} - 
\frac{w}{2}-1 \ .
\end{equation}
It has the limiting behavior
$\pi_{\rm w}(w\to\pm \infty) \simeq -\lambda_{\mp}^*w$, which is
consistent with the exponential tails in (\ref{tail_Pw}).
In terms of the LDF, the DFT for the work is written as 
$\gamma_{\rm w}(\lambda)=\gamma_{\rm w}(1-\lambda)$ and 
$\pi_{\rm w}(w)-\pi_{\rm w}(-w) = -w$. 
The explicit solutions in (\ref{ldf_gw}) and (\ref{ldf_pw}) confirm the DFT.

\subsection{Fluctuations of energy change}
The MGF of the energy change is given by
\begin{equation}\label{Ge_solution}
G_{{\rm e}}(\kappa) = G_{\rm we}(0,\kappa) = B(0,\kappa) \ .
\end{equation} 
It also has simple poles at $\kappa=\kappa_{\pm}(t)$ with $\kappa_\pm^* =
\lim_{t\to\infty}\kappa_\pm(t)=\pm 1$, which are marked with the
closed diamonds in Fig.~\ref{fig1}. The simple poles indicate that 
$P_{\rm e}(E) \sim e^{\kappa_{\mp}(t)E}$ for large $|E|$.
One can obtain $P_{\rm e}(E)$ exactly from the Fourier transform
\begin{equation}
P_{\rm e}(E) = \int_{-\infty}^{\infty} \frac{d\kappa}{2\pi} 
e^{i\kappa E} G_{{\rm e}}(i\kappa) = 
\frac{e^{-|E| / \sqrt{1-e^{-2t}}}}{2\sqrt{1-e^{-2t}}}  \ .
\end{equation}
It indeed decays exponentially. The LDF is given by
\begin{equation}\label{ldf_e}
\pi_{\rm e}(e) = |e| \ .
\end{equation}

The energy change is distributed symmetrically. 
So, the system may gain or lose the energy equally probably. 
It is interesting to note that the PDF is
independent of the driving strength $\varepsilon$. 
The independence is due to a specific property of the force matrix 
in (\ref{F_solvable}). The nonconservative force $\bmf_{nc} = i\varepsilon
\sigma_y \cdot \bmx$ is perpendicular to the conservative force
$\bmf_c = -\bm{\nabla}V(\bmx) = -\bmx$. Hence the nonequilibrium force 
does not affect the energy fluctuation. 

\subsection{Fluctuations of heat}
\begin{figure}
\includegraphics*[width=\columnwidth]{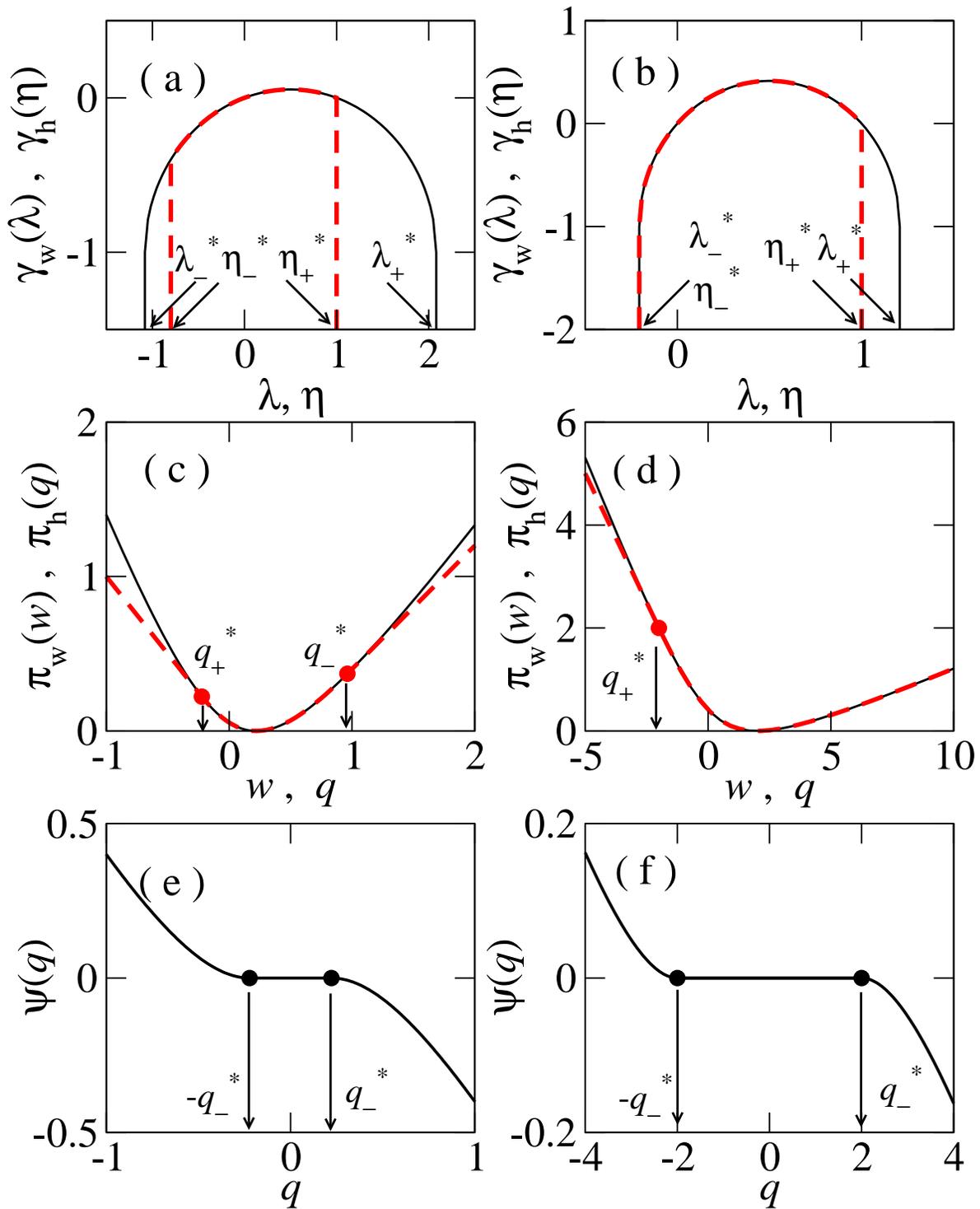}
\caption{(Color online) LDFs for the MGFs in (a) and (b) and for 
the PDFs in (c) and (d). 
Solid and dashed lines are for the work and the heat, respectively.
The panels (e) and (f) show the plot of $\psi(q)$.
The parameter values are $\varepsilon = 1/3$ 
in (a), (c), and (e), and $\varepsilon=1$ in (b), (d), and (f).} 
\label{fig2}
\end{figure}

The MGF of the heat is given by 
\begin{equation}\label{Geta}
G_{{\rm h}}(\eta) = G_{\rm he}(\eta,\kappa=0) = B(\eta,-\eta) \ .
\end{equation}
It has simple poles at $\eta=\eta_{\pm}(t)$, which are
marked with closed squares in Fig.~\ref{fig1}.
From the $t\to\infty$ limit of (\ref{Geta}), the LDF 
is given by 
\begin{equation}
\gamma_{\rm{\rm h}}(\eta) = \left\{
\begin{array}{ccc}
\Omega(\eta)-1 &,& \eta_-^* < \eta < \eta_+^* \\ [2mm]
-\infty &,& \mbox{otherwise}
\end{array}\right.
\end{equation}
where
\begin{equation}
\eta_+^*  =  1 
\end{equation}
and 
\begin{equation}
\eta_-^* = \left\{ \begin{array}{ccc}
      - (1-\varepsilon^2)/(1+\varepsilon^2) &,& \varepsilon^2\leq 1/3 \\
        [2mm]
      - \frac{1}{2}\left(\sqrt{1+1/\varepsilon^2}-1\right) &,&
        \varepsilon^2>1/3 
      \end{array}\right.
\end{equation}
Note that
$\gamma_{{\rm h}}(\eta)$ has the same function form as 
$\gamma_{{\rm w}}(\lambda)$, but $\gamma_h$ is supported in the narrower
domain. We compare the two LDFs in Fig.~\ref{fig2}(a) and (b). 

It is straightforward to obtain the LDF $\pi_{\rm h}(q)$ 
from the Legendre transformation 
$\pi_{\rm h}(q) = \max_\eta\{ \gamma_{{\rm h}}(\eta) - \eta q\}$: 
\begin{equation}\label{ldf_q}
\pi_{\rm h}(q) = \left\{ \begin{array}{ccl}
- q &,& q < q_+^* \\ [2mm]
\sqrt{\frac{(1+\varepsilon^2)(q^2+4\varepsilon^2)}
{4\varepsilon^2}}-\frac{q}{2}-1 &,& q_+^* \leq q \leq q_-^* \\ [2mm]
\frac{1-\varepsilon^2}{1+\varepsilon^2} q -
\frac{4\varepsilon^2}{1+\varepsilon^2} &,& q_-^* < q 
\end{array}\right.
\end{equation}
where 
\begin{equation}\label{q+*}
q_+^* = \left. \frac{d\gamma_{\rm h}(\eta)}{d\eta}\right|_{\eta=\eta_+^*} = -2\varepsilon^2
\end{equation} 
and 
\begin{equation}\label{q-*}
q_-^* = \left. \frac{d\gamma_{\rm h}(\eta)}{d\eta}\right|_{\eta=\eta_-^*} = 
\left\{ \begin{array}{ccl}
2\varepsilon^2 \frac{(3 - \varepsilon^2)}{(1-3\varepsilon^2)} &,& \varepsilon^2<1/3 \\ 
\infty &,& \varepsilon^2\geq 1/3 \ .
\end{array}\right.
\end{equation}
The LDF for the heat is compared with that
of the work in Fig.~\ref{fig2}(c) and (d). They deviate from each other at
large values of $|w|$ and $|q|$. The heat exhibits
stronger fluctuations than the work.

We test whether the LDF of the heat satisfies the DFT. 
The modified fluctuation relation in (\ref{DFT_Q}) is rewritten as
\begin{equation}\label{DFT_ldf_q}
\pi_{\rm h}(q)-\pi_{\rm h}(-q) = -q-\psi(q)
\end{equation}
where
\begin{equation}
\psi(q) \equiv -\lim_{t\to\infty} \frac{1}{t} \ln \Psi(qt)
\end{equation}
with $\Psi(Q)$ in (\ref{Psi_def}). 
Using (\ref{ldf_q}), we find that
\begin{equation}\label{psi_res}
\psi(q) = \left\{ \begin{array}{ccc}
   0 &,& 0 \leq q < -q_+^* \\ [2mm]
   \frac{q}{2} - 
      \frac{\sqrt{(1+\varepsilon^2)(q^2+4\varepsilon^2)}}{2\varepsilon}+1 
       &,& -q_+^* \leq q < q_-^*  \\ [2mm]
   -\frac{1-\varepsilon^2}{1+\varepsilon^2}q + 
       \frac{4\varepsilon^2}{1+\varepsilon^2} &,& q_-^*\leq q 
\end{array}\right.
\end{equation}
and $\psi(-q)=-\psi(q)$. It is plotted in Fig.~\ref{fig2}(e) and (f).
The heat appears to obey the DFT with $\psi(q)=0$ 
within the interval $|q| \leq |q_+^*|$. However, rare fluctuations with
large values of $|q|$ do not obey the fluctuation theorem.

\subsection{Correlations between thermodynamic quantities}
We have shown that the heat does not obey the DFT even in the $t\to\infty$
limit. The correction factor $\psi(q)$ calculated in (\ref{psi_res}) does not vanish
in the large $|q|$ region. Note that $\Psi(Q)\sim e^{-t\psi(Q/t)}$ 
reflects the mutual correlation between the heat and the
energy change in nonequilibrium dynamics. In this subsection, we investigate
the mutual correlations among them.

The LDF for the joint distribution of $\calW$ and $\Delta\calE$ is defined by 
\begin{eqnarray}
\gamma_{\rm we}(\lambda,\kappa) &\equiv& - \lim_{t\to\infty} \frac{1}{t}\ln
G_{\rm{we}}(\lambda,\kappa) \ ,  \\
\pi_{\rm we}(w,e) &=& -\lim_{t\to\infty} \frac{1}{t} P_{\rm we}(W=wt,E=et) \
.
\end{eqnarray}
Using the expression in (\ref{Gwe_lin_sol}) and the analytic
property of the function $B$ defined in (\ref{B_def}), we obtain that
\begin{equation}
\gamma_{\rm we}(\lambda,\kappa) = \left\{ \begin{array}{ccc}
\Omega(\lambda) - 1 &,& (\lambda,\kappa) \in \calD_{\rm we} \\ [2mm]
-\infty &,& \mbox{otherwise}\end{array}\right.
\end{equation}
where $\calD_{\rm we} \equiv \left\{ (\lambda,\kappa) |
\lambda_-^* < \lambda < \lambda_+^*, | \kappa | <
\frac{\Omega(\lambda)+1}{2} \right\}$ denotes the domain 
bounded by the dashed line in Fig.~\ref{fig1}.
The Legendre transformation
\begin{equation}
\pi_{\rm we}(w,e) = \max_{(\lambda,\kappa)\in\calD_{\rm we}} 
\{ \gamma_{\rm we}(\lambda,\kappa) - \lambda w - \kappa e\}
\end{equation}
yields that
\begin{equation}\label{pi_we}
\pi_{\rm we}(w,e) =
\sqrt{\frac{(1+\varepsilon^2)(w^2+\varepsilon^2(2+|e|)^2)}{4\varepsilon^2}} - 
\frac{w-|e|}{2}-1 \ .
\end{equation}
It is minimum at $(w,e)=(2\varepsilon^2,0)$ and increases linearly in $|q|$
and $|e|$ asymptotically.

The joint distribution of $\calQ$ and $\Delta\calE$ is related to that
of $\calW$ and $\Delta\calE$ through (\ref{he_we}) and (\ref{G_he}). Hence,
the LDFs $\gamma_{\rm he}(\eta,\kappa)$ for $G_{\rm he}(\eta,\kappa)$ 
and $\pi_{\rm he}(q,e)$ for $P_{\rm he}(Q,E)$ are given by
\begin{eqnarray}
\gamma_{\rm he}(\eta,\kappa) &=& \gamma_{\rm we}(\eta,\kappa-\eta) \ ,
\label{g_he} \\
\pi_{\rm he}(q,e) &=& \pi_{\rm we}(q+e,e) \ . \label{pi_he}
\end{eqnarray}

\begin{figure}[t]
\includegraphics*[width=\columnwidth]{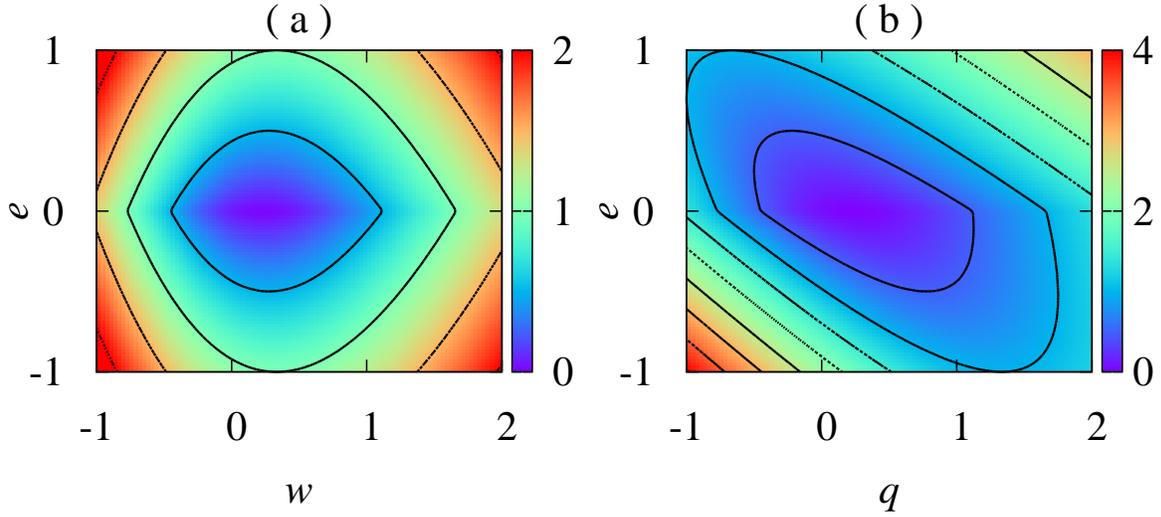}
\caption{(Color online) Density plots for the LDFs $\pi_{\rm we}(w,e)$ in (a)
and $\pi_{\rm he}(q,e)$ in (b) at $\varepsilon=1/3$. Also drawn are the
equiprobable lines.}\label{fig3}
\end{figure}

We compare $\pi_{\rm we}(w,e)$ and $\pi_{\rm he}(q,e)$
at $\varepsilon=1/3$ with the density plot in Fig.~\ref{fig3}. 
Interestingly, $\pi_{\rm we}(w,e)=\pi_{\rm we}(w,-e)$ at all values of $w$. 
Irrespective of an external work, the internal energy of the system may 
increase or decrease with the equal probability. By contrast, 
a strong anti-correlation exists between the heat dissipation and the energy
change. The system tends to lose an internal energy when it dissipates 
a heat and to gain an internal energy when it absorbs a heat.

The mutual correlation is represented well by the most probable value of the
energy change $e_{pr}(q)$ to a given value of $q$. 
That is to say, $\pi_{\rm he}(q,e)$ is minimum at $e=e_{pr}(q)$ when it is
regarded as a function of $e$ with fixed $q$. Using the explicit solution
for $\pi_{\rm he}(q,e)$ in (\ref{pi_we}) and (\ref{pi_he}), we find that
\begin{equation}\label{e_pr_res}
e_{pr}(q) = \left\{ \begin{array}{ccl}
 -\frac{q+2\varepsilon^2}{1+\varepsilon^2} &,& q< q_+^* \\ [2mm]
 0  &,& q_+^* \leq q < q_-^* \\ [2mm]
 \frac{-(1-3\varepsilon^2)q+2\varepsilon^2(3-\varepsilon^2)}{1-\varepsilon^4}
&,& q>q_-^*
\end{array}\right.
\end{equation}
It takes a positive value for $q<q_+^*~(<0)$ and a negative value for
$q>q_-^*~(>0)$. 

We will show that the nonzero $e_{pr}(q)$ is directly related to the
breakdown of the DFT for the heat. The DFT can be examined by checking
whether the relation $\pi_{\rm h}(q)-\pi_{\rm h}(-q)=-q$ holds or not.
Note that $P_{\rm h}(Q) = \int dE P_{\rm
he}(Q,E) \sim \int de\ \exp[-t\pi_{\rm he}(Q/t,e)]\sim \exp[-t \min_{e}
\{\pi_{\rm he}(Q/t,e)\}]$, which yields that 
$\pi_{\rm h}(q) = \pi_{\rm he}(q,e_{pr}(q))$. Then, it follows that
\begin{eqnarray*}
\pi_{\rm h}(q)-\pi_{\rm h}(-q) &=& \pi_{\rm he}(q,e_{pr}(q))-\pi_{\rm
he}(-q,e_{pr}(-q)) \\
&=& \pi_{\rm he}(q,0)-\pi_{\rm he}(-q,0) \\
&&+ \left[\pi_{\rm he}(q,e_{pr}(q))-\pi_{\rm he}(q,0)\right] \\
&&- \left[\pi_{\rm he}(-q,e_{pr}(-q))-\pi_{\rm he}(-q,0)\right]  \ .
\end{eqnarray*}
Comparing this relation with (\ref{DFT_ldf_q}) and using 
$\pi_{\rm he}(q,0)-\pi_{\rm he}(-q,0) =
\pi_{\rm we}(q,0)-\pi_{\rm wr}(-q,0)=-q$, we obtain an alternative
expression for $\psi(q)$:
\begin{eqnarray}
\psi(q) &=& \left[\pi_{\rm he}(q,e_{pr}(q))-\pi_{\rm he}(q,0)\right] 
            \nonumber \\
        &-& \left[\pi_{\rm he}(-q,e_{pr}(-q))-\pi_{\rm he}(-q,0)\right] \ .
\end{eqnarray}
It would vanish if $e_{pr}(q)$ were equal to 0 at all $q$. The
negative correlation between the heat and the energy change leads to nonzero
values of $e_{pr}(q)$, hence nonzero values of $\psi(q)$. 
This analysis shows that the correlation is the origin of the breakdown of
the DFT for the heat.

\section{Summary and discussions}\label{sec5}
We have derived the modified fluctuation relation for the heat given in
(\ref{DFT_Q}) in general overdamped Langevin systems. It involves the extra
factor $\Psi(Q)$ defined in (\ref{Psi_def}) which reflects the correlation
between the dissipated heat and the energy change of the system. 
We have investigated the mutual correlations in a two-dimensional linear
diffusion system, which is exactly solvable by using the method presented 
in~\ref{appendix}. We have obtained the closed form expressions for the
LDFs for the probability distributions in (\ref{ldf_pw}),
(\ref{ldf_e}), (\ref{ldf_q}), (\ref{pi_we}), and (\ref{pi_he}). 
In particular, the result shows that the heat and the energy change
are negatively correlated. It is manifested in $e_{pr}(q)$ that corresponds
to the value of $e$ minimizing $\pi_{\rm he}(q,e)$ with fixed $q$, i.e., 
the most probable energy change to a given heat dissipation. 
It is nonzero for large values of $q$ as shown in (\ref{e_pr_res}), and
gives rise to a nonzero value of $\psi(q)$. Therefore, we conclude
that the negative correlation is the origin for the distinct fluctuation
property of the heat.

In this work, we only study the simplest linear diffusion system with the
force matrix given in (\ref{F_solvable}). 
A natural direction for future works is to consider a general force matrix
having different diagonal elements, which can describe experimental
systems such as a coupled RC circuit investigated in Ref.~\cite{Ciliberto13}. 
It will be interesting to study the correlations in the anisotropic system. 
It will be also interesting to study the correlations in nonlinear systems. 
We hope that our work triggers further theoretical and experimental studies.

\ack
We would like to thank Hyunggyu Park, Chulan Park, and Jong-Min Park for
helpful discussions. This work was supported by the Basic Science Research
Program through the NRF Grant No.~2013R1A2A2A05006776.

\appendix
\section{Joint distributions in linear diffusion systems}\label{appendix}
In this Appendix, we introduce a path integral formalism for the MGF $G_{\rm
we}(\lambda,\kappa)$ in a linear
diffusion system where the force is given by the form in (\ref{F_lin}). 
It is an extension of the path integral formalism for $G_{\rm
w}(\lambda)$ developed in Ref.~\cite{Kwon11}.

Suppose that the initial configuration $\bmx(t_i)$ 
follows the distribution given by
\begin{equation}\label{P0}
P_i(\bmx) = \sqrt{\det\left(\frac{\sfS}{2\pi} \right)} 
e^{-\frac{1}{2}\bmx^T \cdot \sfS \cdot \bmx} 
\end{equation}
with a symmetric positive-definite matrix $\sfS$. 
The equilibrium Boltzmann distribution is obtained by taking $\sfS=\sfF_s$.
In general, it can be taken arbitrarily depending on a physical condition.
The MGF $G_{\rm w}(\lambda)$ is given by~\cite{Kwon11}
\begin{equation}\label{Gw}
G_{\rm w}(\lambda) = \int[{D}\bmx(t)] \calT[\bmx(t)|\bmx(t_i)]
P_i(\bmx(t_i)) e^{-\lambda \calW[\bmx(t)]}  \ ,
\end{equation}
where $\calT[\bmx(t)|\bmx(t_i)]$ is the conditional probability for a path
$\bmx(t)$ to a given initial configuration $\bmx(t_i)$ and the
work is given in (\ref{LDS_W}). The Onsager-Machlup
formalism~\cite{Onsager53,Lau07} allows one to write the conditional probability
as
\begin{equation}\label{path_prob}
\calT[\bmx(t)|\bmx(t_i)] \propto e^{-\frac{1}{4}\int_{t_i}^{t_f} dt |
\dot\bmx(t)-\bmf(\bmx(t))|^2 - \frac{1}{2}\bm\nabla_{\bmx} \cdot \bmf(t)}
\end{equation}
with the Stratonovich integral.
Note that the exponents in (\ref{LDS_W}), (\ref{P0}), and (\ref{path_prob})
are quadratic in $\bmx$. Hence, $G_{\rm w}(\lambda)$ in (\ref{Gw}) can be
evaluated by the Gaussian integration in principle.

The Gaussian integration can be evaluated efficiently. 
First, discretize the time and introduce $\{\bmx_0, \bmx_1, \cdots, \bmx_M\}$ 
with $\bmx_k=\bmx(t_k=t_i+k(t_f-t_i)/M)$ representing the configuration 
at $k$-th time slice. Then, the MFT in (\ref{Gw}) is written in the form of 
$$
G_{\rm w}(\lambda) \propto \int \left[d\bmx_M \prod_{k=0}^{M-1} d\bmx_k 
       e^{-K_k (\bmx_{k+1},\bmx_k)} \right] e^{-\bmx_0^T \cdot \sfS\cdot
       \bmx_0} \ , 
$$
where $K_k(\bmx_{k+1},\bmx_{k})$ is quadratic in both $\bmx_x$ and
$\bmx_{k+1}$ that is determined from the expressions for $\calT$ and
$\calW$~\cite{Kwon11}. 
The kernel for $\bmx_0$ is $\sfS$.
Since $\bmx_0$ is coupled to $\bmx_1$ through $K_0$, one obtains 
an effective kernel for $\bmx_1$ after integrating over $\bmx_0$. 
It is done successively to obtain a recursion relation for the kernel, 
denoted by $\tilde\sfA(t_k)$, for $\bmx(t_k)$. 
In the $M\to\infty$ limit, the recursion relation can be casted into 
the differential equation
\begin{equation}\label{dAdt}
\frac{d\tilde\sfA}{dt} = -2 \tilde\sfA^2 + \tilde\sfA\tilde\sfF +
\tilde\sfF^T \tilde\sfA + \mathsf{\Lambda} \ ,
\end{equation}
where $\tilde\sfF = \sfF - \lambda (\sfF - \sfF^T)$ and $\mathsf{\Lambda} =
(\sfF^T \sfF - \tilde\sfF^T \tilde \sfF)/2$. The initial condition is
given by 
\begin{equation}\label{dAdt_initial}
\tilde\sfA(t_i) = \sfS \ .
\end{equation} 
Collecting all the factors coming from the all integrations, one obtains
that 
\begin{equation}\label{Gw_formal}
G_{\rm w}(\lambda) =
\sqrt{\frac{\det(\sfS)}{\det(\tilde\sfA(t_f))}}\ e^{-\int_{t_i}^{t_f}dt
\ {\rm Tr}(\tilde\sfA(t)-\tilde\sfF) } \ .
\end{equation}
The factor $\det(\sfS)$ originates from the normalization factor of $P_i$ in
(\ref{P0}) and the factor $\det(\tilde\sfA(t_f))$ from the final Gaussian
integration over $\bmx(t_f)$. The exponential factor accounts for the 
contribution from the intermediate time steps~\cite{Kwon11}.

The formal expression for $G_{\rm we}(\lambda,\eta)$ is given by
\begin{eqnarray}\label{Gwe}
G_{\rm we}(\lambda,\kappa) &=& \int [D\bmx(t)] \calT[\bmx(t)|\bmx(t_i)]
P_i(\bmx(t_i)) \nonumber \\
&\times& e^{-\lambda\calW[\bmx(t)]-\kappa \Delta\calE[\bmx(t))]}
\end{eqnarray}
Note that $\Delta \calE[\bmx]$ is also quadratic in $\bmx$. Hence, $G_{\rm
we}$ can be evaluated using the same method.
Comparing the integrands in (\ref{Gw}) and (\ref{Gwe}), 
one finds that they differ by the boundary term $-\kappa \Delta\calE[\bmx(t)] =
-\frac{\kappa}{2} \bmx(t_f)^T \cdot \sfF_s \cdot \bmx(t_f) + \frac{\kappa}{2}
\bmx(t_i)^T \cdot \sfF_s \cdot \bmx(t_i)$ in the exponent. Consequently,
$G_{\rm{we}}(\lambda,\kappa)$ is simply obtained by replacing
$\sfS$ in (\ref{dAdt_initial}) with $(\sfS-\kappa \sfF_s)$ and
$\tilde\sfA(t_f)$ in (\ref{Gw_formal}) with
$(\tilde\sfA(t_f)+\kappa \sfF_s)$. Therefore, we obtain 
\begin{equation}\label{Gwe_formal}
G_{\rm{we}}(\lambda,\kappa) = 
\sqrt{\frac{\det(\sfS)}{\det(\tilde\sfA(t_f)+\kappa \sfF_s)}}\ 
e^{-\int_{t_i}^{t_f} dt \ {\rm Tr}(\tilde\sfA(t)-\tilde\sfF) } \ ,
\end{equation}
where $\tilde\sfA(t)$ is the solution of (\ref{dAdt}) with the
shifted initial condition
\begin{equation}
\tilde\sfA(t_i) = \sfS - \kappa \sfF_s \ .
\end{equation}
The initial state of the system at time $t=t_i$ is characterized by $\sfS$. 
If one takes the equilibrium Boltzmann distribution 
as the initial state, it should be taken as
\begin{equation}
\sfS = \sfF_s \ .
\end{equation}


\section*{References}
\begin{thebibliography}{99}

\bibitem{Evans93} D.J. Evans, E.G.D. Cohen, and G.P. Morriss, 
        Phys. Rev. Lett. {\bf 71}, 2401 (1993).
\bibitem{Gallavotti95} G. Gallavotti and E.G.D. Cohen, 
        Phys. Rev. Lett. {\bf 74}, 2694 (1995).
\bibitem{Jarzynski97} C. Jarzynski, Phys. Rev. Lett. {\bf 78}, 2690 (1997).
\bibitem{Crooks99} G.E. Crooks, Phys. Rev. E {\bf 60}, 2721 (1999).
\bibitem{Kurchan98} J. Kurchan, J. Phys. A {\bf 31}, 3719 (1998).
\bibitem{Lebowitz99} J.L. Lebowitz and H. Spohn, 
        J. Stat. Phys. {\bf 95}, 333 (1999).
\bibitem{Hatano01}    T. Hatano and S.-i. Sasa,
        Phys. Rev. Lett. {\bf 86}, 3463 (2001).
\bibitem{Seifert05}  U. Seifert, Phys. Rev. Lett. {\bf 95}, 040602 (2005).
\bibitem{Esposito10} M. Esposito and C. Van den Broeck, 
        Phys. Rev. Lett. {\bf 104}, 090601 (2010).
\bibitem{Sagawa10} T. Sagawa and M. Ueda, 
        Phys. Rev. Lett. {\bf 104}, 090602 (2010).
\bibitem{Spinney12} R.E. Spinney and I.J. Ford, 
        Phys. Rev. Lett. {\bf 108}, 170603 (2012).
\bibitem{Lee13} H.K. Lee, C. Kwon, and H. Park, 
        Phys. Rev. Lett. {\bf 110}, 050602 (2013).
\bibitem{Wang02} G.M. Wang, E.M. Sevick, E. Mittag, D.J. Searles, and D.
        J. Evans, Phys. Rev. Lett. {\bf 89}, 050601 (2002).
\bibitem{Hummer01} G. Hummer, A. Szabo, 
        Proc. Natl. Acad. Sci. USA {\bf 98}, 3658 (2001).
\bibitem{Liphardt02} J. Liphardt, S. Dumont, S.B. Smith, I. Tinoco Jr., 
        and C. Bustamante, Science {\bf 296}, 1832 (2002).
\bibitem{Hayashi10} K. Hayashi, H. Ueno, R. Iino, and H. Noji,
        Phys. Rev.  Lett. {\bf 104}, 218103 (2010).
\bibitem{Ciliberto10} S. Ciliberto, S. Joubaud, and A. Petrosyan, 
        J. Stat. Mech. (2010) P12003.
\bibitem{Seifert12} U. Seifert, Rep. Prog. Phys. {\bf 75}, 126001 (2012).
\bibitem{Garcia10} R. Garc\'ia-Garc\'ia, D. Dom\'inguez, V. Lecomte, and 
        A.B. Kolton, Phys. Rev. E {\bf 82}, 030104(R) (2010).
\bibitem{Garcia12} R. Garc\'ia-Garc\'ia, V. Lecomte, A. B. Kolton, and 
        D. Dom\'inguez, J. Stat. Mech. (2012) P02009.
\bibitem{Noh12} J.D. Noh and J.-M. Park, 
        Phys. Rev. Lett. {\bf 108}, 240603 (2012).
\bibitem{Farago02} J. Farago, J. Stat. Phys. {\bf 107}, 781 (2002).
\bibitem{Zon03} R. van Zon and E.G.D. Cohen, 
        Phys. Rev. Lett. {\bf 91}, 110601 (2003).
\bibitem{Zon04} R. van Zon and E.G.D. Cohen, 
        Phys. Rev E {\bf 69}, 056121 (2004).
\bibitem{Garnier05} N. Garnier and S. Ciliberto,
        Phys. Rev. E {\bf 71}, 060101(R) (2005).
\bibitem{Visco06} P. Visco, J. Stat. Mech. (2006) P06006.
\bibitem{Harris06} R.J. Harris, A. R\'akos, and G.M. Sch\"utz, 
        Europhys. Lett. {\bf 75}, 227 (2006).
\bibitem{Rakos08} A. R\'akos and R.J. Harris, 
        J. Stat. Mech. (2008) P05005.
\bibitem{Fogedby11} H.C. Fogedby and A. Imparato,
        J. Stat. Mech. (2011) P05015.
\bibitem{Nemoto12} T. Nemoto, Phys. Rev. E {\bf 85}, 061124 (2012).
\bibitem{Puglisi06} A. Puglisi, L. Rondoni, and A. Vulpiani,
        J. Stat. Mech. (2006) P08010.
\bibitem{JLee13} J.S. Lee, C. Kwon, and H. Park, 
        Phys. Rev. E {\bf 87}, 020104(R) (2013).
\bibitem{Kwon11} C. Kwon, J.D. Noh, and H. Park,
        Phys. Rev. E {\bf 83}, 061145 (2011).
\bibitem{Noh13} J.D. Noh, C. Kwon, and H. Park,
        Phys. Rev. Lett. {\bf 111}, 130601 (2013).
\bibitem{Kwon13} C. Kwon, J.D. Noh, and H. Park,
        Phys. Rev. E {\bf 88}, 062102 (2013). 
\bibitem{gardiner} C.W. Gardiner, {\em Handbook of Stochastic Methods for
Physics, Chemistry and the Natural Sciences}, 2nd ed. (Springer, Berlin,
1985).
\bibitem{risken} H. Risken, {\it The Fokker-Planck Equation}, 2nd ed.
(Springer, Berlin, 1989).
\bibitem{Sekimoto98} K. Sekimoto, 
        Prog. Theor. Phys. Suppl. {\bf 130}, 17 (1998).
\bibitem{Cover06} T. M. Cover and J. A. Thomas,
        {\em Elements of Information Theory} 
        (Wiley, New York, 1991).
\bibitem{Onsager53} L. Onsager and S. Machlup, 
        Phys. Rev. {\bf 91}, 1505 (1953).
\bibitem{comment1} In general, one may decompose the force matrix as $\sfF =
(\sfF_s + \mathsf{O})+(\sfF_a-\mathsf{O})$  with an {\em arbitrary
symmetric} matrix $\mathsf{O}$. With any choice of $\mathsf{O}$, one can
define the energy and the nonequlibrium work consistently as was done in
Ref.~\cite{Kwon11}. In this work, we only consider the specific but natural 
choice of $\mathsf{O}=0$ in order to avoid an unnecessary complication.
\bibitem{Kundu11} A. Kundu, S. Sabhapandit, and A. Dhar, 
        J. Stat.  Mech. (2011) P03007.
\bibitem{Saito11} K. Saito and A. Dhar, Phys. Rev. E {\bf 83}, 041121 (2011).
\bibitem{Sabhapandit12} S. Sabhapandit, Phys. Rev. E {\bf 85}, 021108 (2012).
\bibitem{Fogedby12} H.C. Fogedby and A. Imparato, 
        J. Stat. Mech. (2012) P04005.
\bibitem{Ciliberto13} S. Ciliberto, A. Imparato, A. Naert, and M. Tanase, 
        Phys. Rev. Lett.  {\bf 110}, 180601 (2013).
\bibitem{Touchette09} H. Touchette, Phys. Rep. {\bf 478}, 1 (2009).
\bibitem{Lau07} A.W.C. Lau and T.C. Lubensky, 
        Phys. Rev. E {\bf 76}, 011123 (2007).
\end{thebibliography}
\end{document}